# Fermion-boson many-body interplay in a frustrated kagome paramagnet

**Authors:** J.-X. Yin[1]*†, Nana Shumiya[1]*, Sougata Mardanya[2,11]*, Qi Wang[3]*, S. S. Zhang[1]*, Hung-Ju Tien[2]*, Daniel Multer[1], Yuxiao Jiang[1], Guangming Cheng[4], Nan Yao[4], Shangfei Wu[5], Desheng Wu[5], Liangzi Deng[6], Zhipeng Ye[7], Rui He[7], Guoqing Chang[1], Zhonghao Liu[8], Kun Jiang[9], Ziqiang Wang[9], Titus Neupert[10], Amit Agarwal[11], Tay-Rong Chang[2], Ching-Wu Chu[6,12], Hechang Lei[3], M. Zahid Hasan[1,12]†

**Affiliations:**

[1]Laboratory for Topological Quantum Matter and Spectroscopy (B7), Department of Physics, Princeton University, Princeton, New Jersey, USA.

[2]Department of Physics, National Cheng Kung University, Tainan, Taiwan.

[3]Department of Physics and Beijing Key Laboratory of Opto-electronic Functional Materials&Micro-nano Devices, Renmin University of China, Beijing, China.

[4]Princeton Institute for Science and Technology of Materials (PRISM), Princeton University, Princeton, New Jersey, USA.

[5]Institute of Physics, Chinese Academy of Sciences, Beijing, China.

[6]Department of Physics and Texas Center for Superconductivity, University of Houston, Houston TX, USA.

[7]Department of Electrical and Computer Engineering, Texas Tech University, Lubbock, Texas, USA.

[8]State Key Laboratory of Functional Materials for Informatics and Center for Excellence in Superconducting Electronics, Shanghai Institute

of Microsystem and Information Technology, Chinese Academy of Sciences, Shanghai, China

[9]Department of Physics, Boston College, Chestnut Hill, Massachusetts, USA.

[10]Department of Physics, University of Zurich, Winterthurerstrasse 190, Zurich, Switzerland.

[11]Department of Physics, Indian Institute of Technology Kanpur, Kanpur, India.

[12]Lawrence Berkeley National Laboratory, Berkeley, California, USA.

†Corresponding author, E-mail: jiaxiny@princeton.edu; mzhasan@princeton.edu;

*These authors contributed equally to this work.

## Abstract

**Kagome-net, appearing in areas of fundamental physics, materials, photonic and cold-atom systems, hosts frustrated fermionic and bosonic excitations[1-3]. However, it is extremely rare to find a system to study both fermionic and bosonic modes to gain insights into their many-body interplay. Here we use state-of-the-art scanning tunneling microscopy and spectroscopy to discover unusual electronic coupling to flat-band phonons in a layered kagome paramagnet. Our results reveal the kagome structure with unprecedented atomic resolution and observe the striking bosonic mode interacting with dispersive kagome electrons near the Fermi surface. At this mode's energy, the fermionic quasi-particle dispersion exhibits a pronounced renormalization, signaling a giant coupling to bosons. Through a combination of self-energy analysis, first-principles calculation, and a lattice vibration model, we present evidence that this mode arises from the geometrically frustrated phonon flat-band, which is the lattice analog of kagome electron flat-band. Our findings provide the first example of kagome bosonic mode (flat-band phonon) in electronic excitations and its strong interaction with fermionic degrees of freedom in kagome-net materials.**

## Introduction

The kagome-net, a pattern of corner-sharing triangular plaquettes, has been a fundamental model platform for exotic states of matter, including quantum spin liquids and topological band structures[1-3]. Recently, the transition metal-based kagome metals[4-13] are emerging as a new class of topological quantum materials to explore the interplay between frustrated lattice geometry, nontrivial band topology, symmetry-breaking order and many-body interaction. A kagome lattice tight-binding model generically features a Dirac crossing and a flat-band, which are the fundamental sources of nontrivial topology and strong correlation. Such topological fermionic structures arising from the correlated 3$d$ electrons in the kagome lattice have been widely reported in several quantum materials[4-13], including $Mn_3Sn$, $Fe_3Sn_2$, $Co_3Sn_2S_2$, $TbMn_6Sn_6$, FeSn and CoSn. In parallel, the band dispersion of bosonic excitations on a kagome lattice also features Dirac crossings and flat-bands, as demonstrated in photonic crystals[14,15]. A question naturally arising when studying kagome lattice electrons is the possibility of a nontrivial many-body interplay between the bosonic kagome lattice phonons and the fermionic quasiparticles.

Such fermion-boson interactions often manifest as a perturbation of the bare band structures at very low energy scales. However, most kagome lattice materials exhibit complicated multi-bands near the Fermi energy[4-12], severely challenging the clear identification of the many-body effect by spectroscopic methods. Among all known kagome lattice materials, the paramagnetic CoSn is recently highlighted as an outstanding kagome topological metal with much cleaner bands and simpler Fermi surface[13], making it an ideal platform to search for the geometrical frustrated fermion-boson interaction. Here we report the discovery of fermion-boson many-body interplay in kagome lattice of CoSn, which arises from the coupling of the phonon flat-band with the kagome electrons, utilizing the low-temperature (T = 4.2K), high energy-resolution ($\Delta E < 0.3$ meV), atomic layer resolved scanning tunneling microscopy.

## Results

**Atomic-scale visualization of kagome lattice.** CoSn has a hexagonal structure (space group P6/mmm) with lattice constants[13,16] $a$ = 5.3 Å and $c$ = 4.4 Å. It consists of a $Co_3Sn$ kagome layer and an $Sn_2$ honeycomb layer (Fig. 1a) with alternating stacking. The side-plane atomically resolved map of the crystal measured by transmission electron microscopy (Fig. 1b) directly demonstrates this atomic stacking sequence along the $c$-axis. Upon cryogenic cleaving, the surface yields either the $Co_3Sn$- or $Sn_2$-terminated atomic layer. The lower panel in Fig. 1c shows a highly rare topographic image of the cleaving surface that contains both terminations. It consists of a $Co_3Sn$ surface and islands of $Sn_2$ layer sitting on top. The simultaneously obtained differential conductance map directly reveals their difference in the electronic structure shown in the upper panel of Fig. 1c, where the $Co_3Sn$ surface has a higher density of states at the bias voltage of 100mV. From this map, we also find that the $Co_3Sn$ surface has detectable impurities induced quasi-particle interferences, which is the basis for our further space-momentum investigation. Scanning the $Sn_2$ (Fig. 1d) and $Co_3Sn$ (Fig. 1e) surfaces with higher magnification, we directly reveal their honeycomb and kagome lattice symmetry, respectively. Remarkably, our topographic image was able to resolve the fine corner-sharing triangle structure of the Co kagome lattice and the Sn atom in the kagome center (Fig. 1e). Such ultra-high atomic resolution has not been achieved in the previous scanning tunneling studies of kagome lattice materials.

**Bosonic mode coupling from kagome electrons.** With the lattice structure of CoSn visualized at the atomic scale, we now study their electronic structure by measuring the differential conductance as shown in Fig. 2a. According to the first principle calculations and photoemission measurement, the Fermi surface is dominated by a fairly simple electron-like band[13]. Strikingly, we find pronounced low-energy modulations for the spectra taken on the $Co_3Sn$ layer, while this feature is absent on the $Sn_2$ layer as shown in Fig. 2b. The observed peak-dip-hump modulation can indicate the strong bosonic mode coupling with the coherent state at the Fermi level, as similar spectroscopic features have been found in many strong coupling superconductors, including lead[17], cuprates[18] and iron-pnictides[19]. Such a bosonic mode often arises from phonons or spin resonances. Since this material is nonmagnetic and has no detectable magnetic field dependence of tunneling spectra up to 8T (Fig. 2c), the bosonic mode is more likely to arise from the coupling to phonons. Moreover, the electronic coupling to the bosonic mode can be described within the Eliashberg theory[20,21] with $\alpha^2 F(\omega)$ where $\alpha$ is the coupling matrix element and $F(\omega)$ is the bosonic density of states, and this can be studied in the tunneling experiments. $\alpha^2 F(\omega)$ is intimately related to the second differentiation of the tunneling spectra[17-22]. Analyzing the measured tunneling spectra (Figs. 2d and e), we find a Gaussian-like state in the second derivative centered at $E_M$ = 15meV with a full width at half maximum of 9meV (Fig. 2e), and identify it as a candidate signal related to the Eliashberg function (Fig. 2e, blue curves). The other peak features at lower energies of the second differential spectra can be expected from a coherent state at the Fermi energy as shown by the simulation curve in Figs. 2d and e.

**Many-body kink in the kagome dispersion.** To gain deeper insight into the bosonic mode coupling on the kagome lattice, we perform systematic spectroscopic imaging on a large $Co_3Sn$ area with only a few $Sn_2$ adatoms (Fig. 3a). By taking the Fourier transform of the differential conductance map (Fig. 3b), we obtain the quasiparticle interference (QPI) data. The QPI data at 0meV (Fig. 3C) shows a clear ring-like signal, consistent with the intra-band scattering of the dominant electron-like Fermi surface[13] centered at Γ. Thus the low-energy QPI dispersion reflects the behavior of the electron-like band crossing the Fermi surface ($Q = 2k$). Analyzing the QPI dispersion along Γ-M direction in Fig. 3d, we observe a pronounced double kink feature, different from its bare band dispersion calculated by the first-principles (dashed line). The spectroscopic kink feature has been identified as a fingerprint of the bosonic mode coupling[23-29] and indicates a giant mode coupling strength. The energy of the QPI kink is around ±15meV, well consistent with the mode energy $E_M$ in the second differential conductance signal. The coupling strength can be estimated from Fermi velocity renormalization $\lambda = v_{f0}/v_f - 1 = 1.8 \pm 0.3$, where $v_{f0}$ and $v_f$ are the Fermi velocities of the bare QPI band and renormalized QPI band, respectively. We also explored the QPI on the $Sn_2$ honeycomb lattice but did not find any clear kink. Hence the unique kagome lattice resolving capability combined with low temperature and high energy-resolution of our advanced spectroscopic technique can be the key for the kink discovery in this material.

**Discussion**
**Self-energy analysis.** The pronounced kink signal from the kagome lattice allows us to analyze the electron many-body self-energy $\Sigma(\omega)$ and further quantify the Eliashberg function $\alpha^2 F(\omega)$. $\Sigma(\omega)$ is intimately related to $\alpha^2 F(\omega)$, the Fermi-Dirac distribution $f(E)$ and the Bose-Einstein distribution $n(\omega)$ (see Supplementary for more details). The real part of the self-energy Re(Σ) is related to the energy difference between the observed kink dispersion and the bare QPI dispersion, while the imaginary part of the self-energy Im(Σ) is related to the electron band broadening, which is inversely proportional to the QPI intensity. Re(Σ) and Im(Σ) are tied to one another through the Kramers-Kronig relation, and we can more accurately acquire Re(Σ) from the QPI data in Fig. 3e. We take the shape of $\alpha^2 F(\omega)$ in reference to the second differential conductance (Fig. 2), and tune the coupling strength $\lambda = \int \alpha^2 F(\omega)/\omega \, d\omega$ and calculate the real part of the self-energy Re(Σ). We find that with $\lambda = 1.9 \pm 0.3$, the calculated Re(Σ) can account for that determined by the experiment (Fig. 3e), which agrees with the estimated λ from Fermi velocity renormalization. We further simulate the QPI signal with this $\alpha^2 F(\omega)$ under the Green function formalism in Fig. 3f, which also shows reasonable consistency with the experimental data both in dispersion and intensity evolution, providing key support to our many-body analysis of the kagome fermion-boson interaction.

**First-principles calculation.** Having characterized the many-body fermion-boson interaction in the kagome lattice, we perform first-principles calculations of the phonon band to understand the nature of the bosonic mode. Firstly, the calculated phonon density of states exhibits a pronounced peak at 15meV (Fig. 4a), coincide with the mode energy in experiments. Secondly, this phonon mode mainly arises from the $Co_3Sn$ kagome layer, consistent with the experiments. Thirdly, the calculation provides momentum space insight into the origin of this peak, in that it arises from a flat-band in phonon momentum space as shown in Fig. 4b. Lastly, through the atomic displacement resolved calculation, we identify that the flat-band phonon is mainly associated with the Co kagome lattice vibrations confined to the line connecting the centers of two neighboring triangles (Fig. 4b inset).

**Lattice vibration model.** In light of the first-principle calculation, we build a kagome lattice vibration model to elucidate the striking physics. The essential momentum features of the flat-band can be well captured by this model (Fig. 4c), with the flat-band toughing a parabolic band bottom. This model is highly analytical and offers a heuristic understanding of the non-propagating nature of the kagome flat-band phonon mode. We find that the collective lattice displacement shown in the inset of Fig. 4c, a deformation from a hexagonal ring (inner six atoms) rotating clockwise or anti-clockwise, would not exert any net force on the outer atoms. Hence such geometrically frustrated vibrations can be localized forming the phonon flat-band. It is also clear now that this phonon flat-band is the lattice analog of kagome electron flat-band[8,13], whose quadratic band touching feature distinguishes them from the isolated flat-bands in heavy-fermion systems[30] and the Dirac cone touching flat-bands in Moire lattices[31]. The coupling to the kagome phonon flat-band is not predicted by existing research papers but is highly anticipated to explain the giant fermion-boson interaction observed here. Our findings suggest that the flat phonon dispersion can be probed by future momentum-resolved phonon-sensitive scattering experiments including inelastic X-ray scattering and neutron scattering.

The correspondence between the kagome lattice, tunneling conductance, magnetic field response, double kink feature, self-energy analysis, first-principles, and lattice vibration model provides strong evidence and conceptual framework for the fermion-boson interaction in a geometrically frustrated topological quantum material. The nontrivial kagome band structures have been widely observed in both fermionic and bosonic systems[15,32], but their many-body interactions were rarely experimentally observed previously. We expect the latter to be quite general in many topological quantum materials with flat-bands. Such fermion-boson interactions can be the driving force for future discovery of incipient density waves and superconductivity in kagome lattice materials through pressure tuning or chemical engineering. While our research addresses the coupling of the dispersive electrons and the flat band phonon, it would be interesting to explore in the future the intriguing possibility of coupling of the kagome flat electron band and flat phonon band when they are tuned to the similar energies.

**Figures and captions**

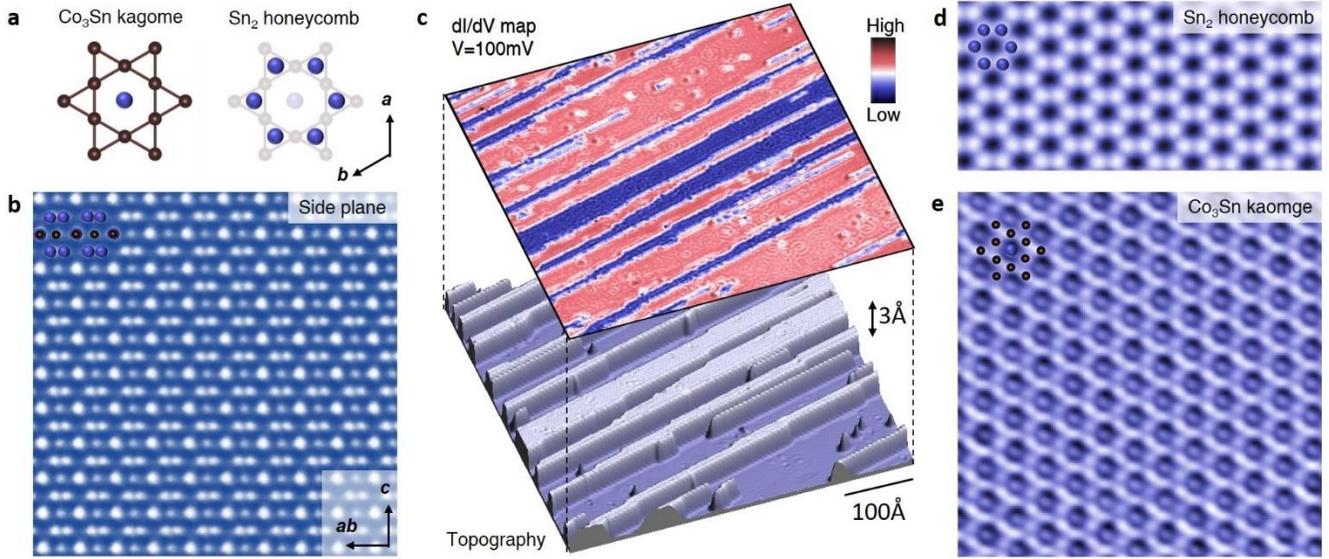

**Fig. 1 Complete atomic-scale visualization of kagome lattice in CoSn. a** Crystal structure of CoSn, which consists of a Co$_3$Sn kagome layer (left) and a Sn$_2$ honeycomb layer (right). **b** Cross-sectional atomic resolution scanning transmission electron microscope image of CoSn, showing the lattice stacking along the *c*-axis as illustrated in the inset. **c** Topographic image of a single atomic step (bottom) and the corresponding differential conductance map obtained at a bias voltage of 100mV (top). **d** Atomically resolved Sn$_2$ surface with the corresponding atomic lattice structures. **e** Atomically resolved Co$_3$Sn surface with the corresponding atomic lattice structures.

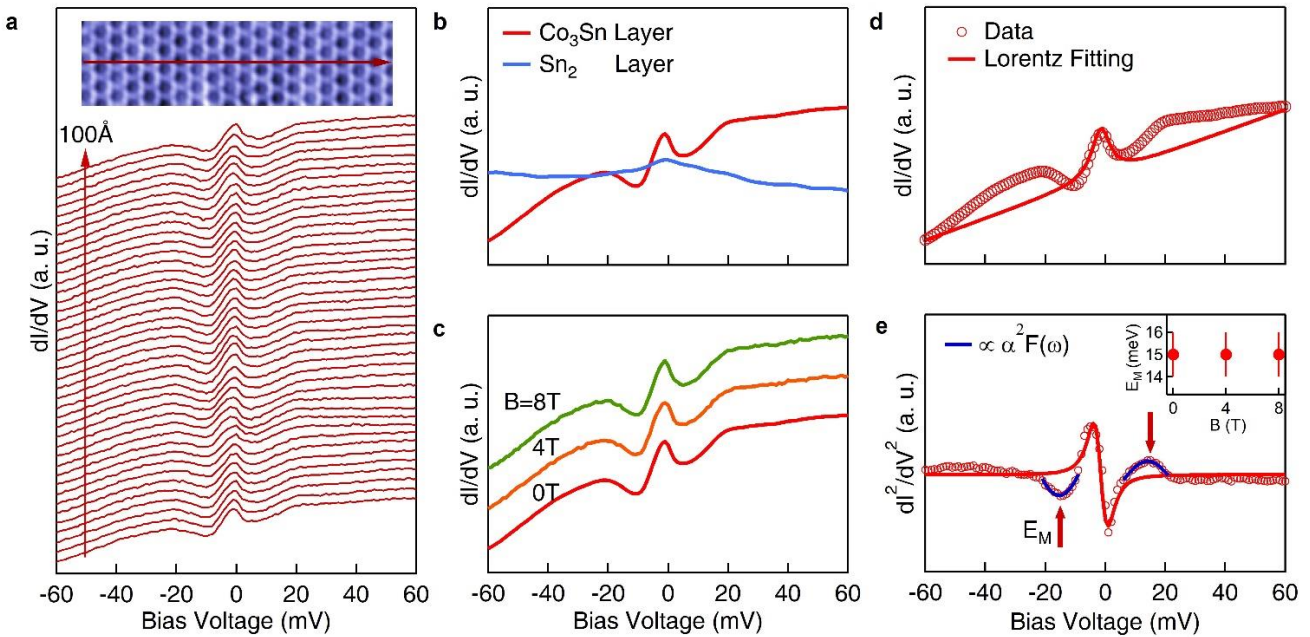

**Fig. 2 Bosonic mode coupling from kagome electrons. a** Differential conductance spectra taken along the Co$_3$Sn layer shown in the inset. **b** Spatially averaged dI/dV spectra for Co$_3$Sn layer and Sn$_2$ layer, respectively. **c** dI/dV spectrum for Co$_3$Sn layer taken at different external magnetic fields applied along the *c*-axis. **d** Fitting of the spectral peak in the Co$_3$Sn layer with a Lorentzian. **e** Second differential

conductance for the data in **d** with the extra Gaussian-like peaks marked by blue lines. The inset plot shows the $E_M$ value is independent of the external magnetic field.

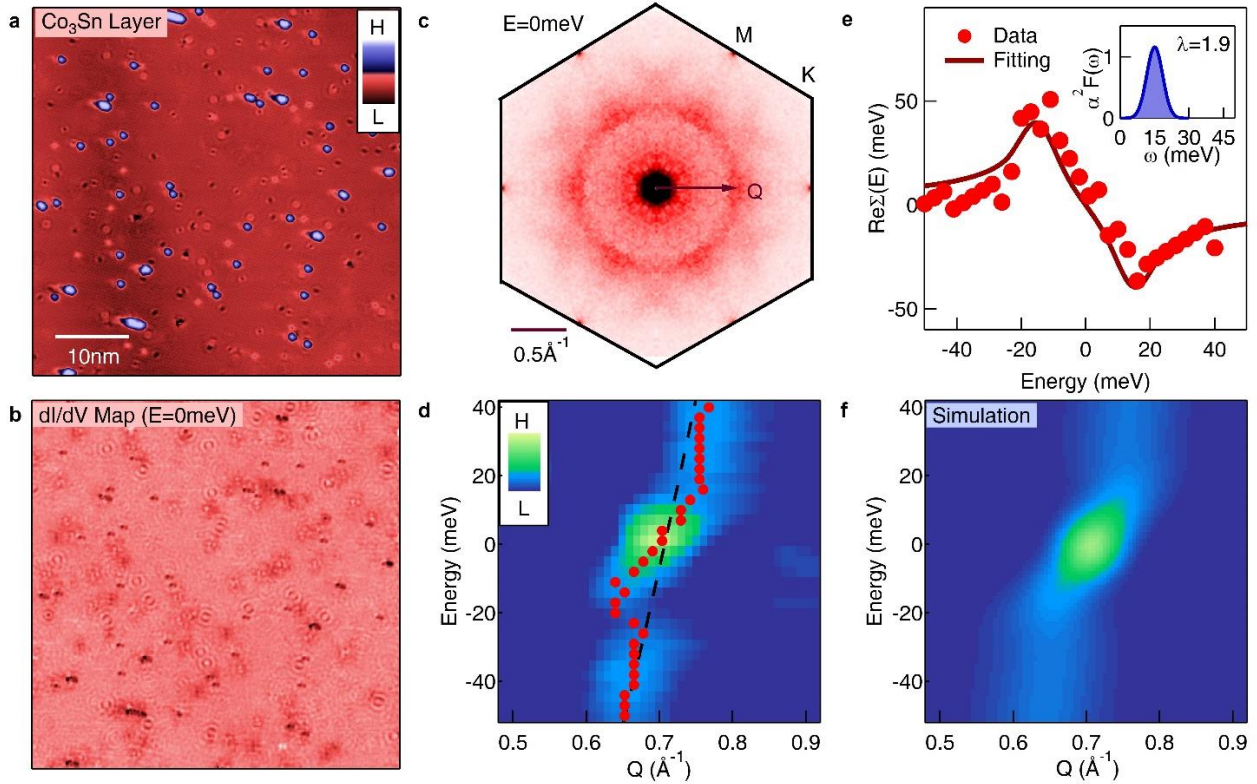

**Fig. 3 Bosonic mode induced many-body kink in the kagome dispersion. a** A topographic image of $Co_3Sn$ layer. **b** Corresponding differential conductance map taking at E = 0meV. **c** QPI data of the $Co_3Sn$ surface. Data has been six-fold symmetrized to enhance the signal to noise ratio. Due to intraband scattering nature, Q = 2k. **d** QPI dispersion along the Γ-M direction showing a pronounced kink. The dots mark the QPI peak positions while the dashed line illustrates the bare band based on first-principles calculations. **e** The real part of the electron self-energy. The dots are extracted from the energy difference between the renormalized QPI dispersion and the bare QPI dispersion. The line is calculated based on a Gaussian-like Eliashberg function with coupling strength λ=1.9±0.3 shown in the inset. **f** Simulated QPI dispersion showing the kink based on the same Eliashberg function.

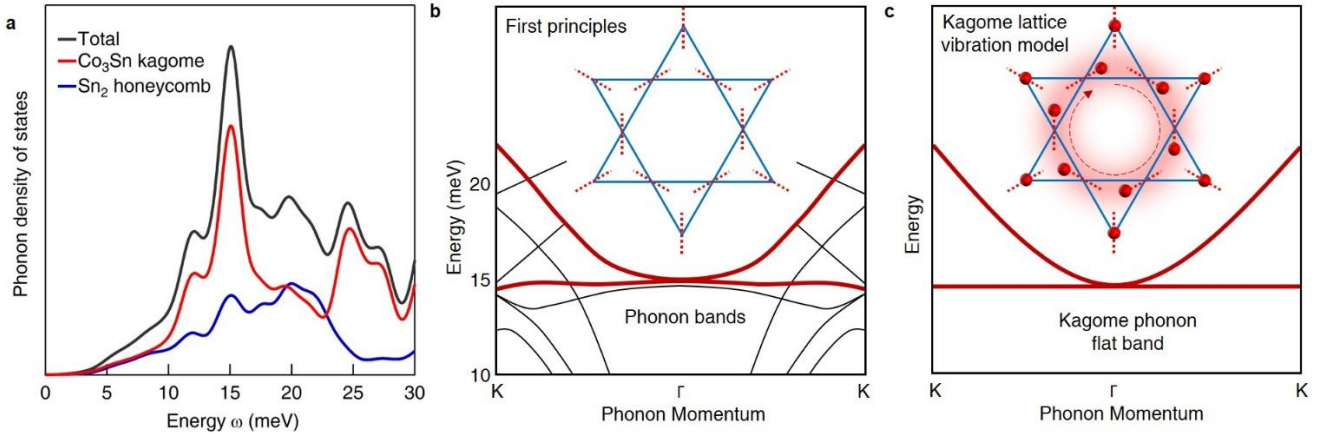

**Fig. 4 Kagome flat-band origin of the bosonic mode. a** Calculated layer-resolved phonon density of states, showing a pronounced peak at 15meV arising from the $Co_3Sn$ kagome layer. **b** Calculated phonon band structure, showing a flat-band at 15meV. The inset shows the calculated atomic displacements for the kagome lattice associated with the phonon flat-band, where the dotted lines denote the atomic vibration directions. **c** Phonon band structure for the kagome lattice vibration model. The inset shows the collective atomic vibrations associated with the phonon flat-band.

## Methods

### Sample preparation

High-quality single crystals of CoSn were synthesized by the Sn flux method. The starting elements of Co (99.99%), Sn (99.99%) were put into an alumina crucible, with a molar ratio of Co : Sn = 3 : 20. The mixture was sealed in a quartz ampoule under a partial argon atmosphere and heated up to 1173 K, then cooled down to 873 K with 2 K/h. The CoSn single crystals were separated from the Sn flux by using a centrifuge.

### Scanning tunneling microscopy characterization

Single crystals with size up to 2mm×2mm×1mm were cleaved mechanically *in situ* at 77K in ultra-high vacuum conditions, and then immediately inserted into the STM head, already at He4 base temperature (4.2K). The magnetic field was applied with zero-field cooling. Tunneling conductance spectra were obtained with an Ir/Pt tip using standard lock-in amplifier techniques with a root mean square oscillation voltage of 0.3meV and a lock-in frequency of 977Hz. Topographic images were taken with tunneling junction set up: V = -100mV I = 2~0.05nA. The conductance maps are taken with tunneling junction set up: V = -50mV, I = 0.5nA.

### Transmission electron microscopy characterization

Thin lamellae were prepared by focused ion beam cutting. All samples for experiments were polished by 2 kV Ga ion beam to minimize the surface damage caused by the high energy ion beam. Transmission electron microscopy imaging, atomic resolution high-angle annular dark-field scanning transmission electron microscopy imaging and atomic-level energy-dispersive X-ray spectroscopy mapping were performed on a commercial scanning/transmission electron microscope equipped with an extreme field emission gun source operated at 300 kV and super-X energy dispersive spectrometry system.

**Many-body theory for bosonic mode coupling**
Due to the scattering by bosonic modes, the electrons do not have a definite energy but rather a finite lifetime and a broadened spectral function. In many-body theory, this phenomenon is characterized by the electron self-energy $\Sigma(\omega)$, which can be regarded as a correction to the free-electron Green function. The electron Green function is

$$G(k,\omega) = \frac{1}{\omega - \epsilon_k^0 - \Sigma(\omega)} \quad (1)$$

where $\epsilon_k^0$ is the electron bare energy dispersion based on first-principle calculation. And the electron spectral function is

$$A(k,\omega) = \frac{1}{\pi}|\text{Im}G(k,\omega)| \quad (2)$$

The QPI dispersion is further described by

$$Q(q,\omega) = -\frac{1}{\pi}\sum_k \text{Im}G(q+k,\omega)G(k,\omega) \quad (3)$$

With the Eliashberg function, the electron self-energy can be described by

$$\Sigma(\omega) = \int dE \int d\omega' \alpha^2 F(\omega') \left[\frac{1-f(E)+n(\omega')}{\omega-\omega'-E} + \frac{f(E)+n(\omega')}{\omega+\omega'-E}\right] \quad (4)$$

where $\alpha^2 F(\omega)$ is Eliashberg coupling function describing the electron-bosonic mode interaction, $f(E)$ is the Fermi-Dirac distribution and $n(\omega)$ is Bose-Einstein distribution at temperature T. Since in most cases, the bosonic mode energy is far less than the Fermi energy $\varepsilon_F$, we can make the approximation that the initial and final energies of scattered electrons are close to Fermi energy. In this way, we obtain an Eliashberg coupling function solely determined by the bosonic energy distribution. An effective coupling constant can be defined as

$$\lambda = 2\int \frac{\alpha^2 F(\omega)}{\omega} d\omega \quad (5)$$

In our experiment, we found a kink in the electron dispersion at $E_M = \pm 15$ meV, which implies a dominant bosonic mode with $\omega_E = 15$ meV. In the calculation, we use the Einstein model and take $\alpha^2 F(\omega)$ as a gaussian peak centered at $\omega_E$. The calculated $\text{Re}\Sigma(k,\omega)$ is given in Fig. 3(E), with comparison to experimental data. With the electron Green function, we can simulate the QPI dispersion $Q(q,\omega)$ by equation (3) convoluted with the experimental resolution, which is given in Fig. 3(F).

**First-principles calculation**
We perform electronic structure calculations within the framework of the density functional theory using norm-conserving pseudopotentials[33] as implemented in the Quantum Espresso simulation package[34]. The exchange-correlation effects are treated within the local density approximation with the Perdew-Zunger parametrization[35]. The electronic calculations used a plane-wave energy cutoff of 80 Ry and a $12 \times 12 \times 12$ $\Gamma$-centered k mesh to sample the Brillouin zone. Total energies were converged to $10^{-7}$ Ry in combination with Methfessel-Paxton type broadening of 0.01 Ry. The phonon calculation is done by using a $2 \times 2 \times 2$ q-mesh grid centered at $\Gamma$ for the sampling of phonon momenta. Starting with the experimental structure, we have fully optimized both the ionic positions and lattice parameters until the Hellmann-Feynman force on each atom is less than $10^{-3}$ Ry/au ($10^{-4}$ Ry) and zero-stress tensors are obtained. We find that the flat-band phonon is mainly associated with the vibrations of Co atoms in $Co_3Sn$ lattice confined to the line connecting the centers of two neighboring triangles as shown by the dotted line in Fig. 5a. Figure 5b shows the atom displacement resolved phonon band structure.

**Kagome lattice vibration model**

We consider a kagome lattice vibration model, assuming the motion of the atoms is highly anisotropic and confined to the dotted line in Fig. 5a. The analysis reproduces the kagome band structure with a flat band quadratically touching another band.

We choose the vectors in real space connecting nearest neighbor kagome atoms as

$$a_1 = (1/2, \sqrt{3}/2) \quad a_2 = (1,0) \quad a_3 = a_1 - a_2 \tag{6}$$

The three atoms $\alpha = 1,2,3$ in each Kagome unit cell can move along the directions

$$u_1 = (\sqrt{3}/2, 1/2) \quad u_2 = (-\sqrt{3}/2, 1/2) \quad u_3 = (0, -1) \tag{7}$$

All atoms are then coupled with the same spring constant, except for the sign. The atoms in one unit cell have the potential energy

$$E_{\text{intra}} = \frac{\lambda}{2} \sum_R \left[ (U_{1,R} - U_{2,R})^2 + (U_{1,R} - U_{3,R})^2 + (U_{2,R} + U_{3,R})^2 \right] \tag{8}$$

In contrast, the inter-unit cell coupling reads

$$E_{\text{inter}} = \frac{\lambda}{4} \sum_R \left[ (U_{1,R} - U_{2,R-2a_2})^2 + (U_{1,R} - U_{3,R-2a_1})^2 + (U_{2,R} + U_{3,R-2a_3})^2 + (U_{2,R} - U_{1,R+2a_2})^2 + (U_{3,R} - U_{1,R+2a_1})^2 + (U_{3,R} + U_{2,R+2a_3})^2 \right] \tag{9}$$

After Fourier transformation, the potential energy reads

$$E_{\text{pot}} = \sum_k (U_{1,k}, U_{2,k}, U_{3,k}) v_k \begin{pmatrix} U_{1,-k} \\ U_{2,-k} \\ U_{3,-k} \end{pmatrix} \tag{10}$$

with

$$v_k = \frac{\lambda}{2} \begin{pmatrix} 4 & -1 - e^{2ik \cdot a_2} & -1 - e^{2ik \cdot a_1} \\ -1 - e^{-2ik \cdot a_2} & 4 & 1 + e^{2ik \cdot a_3} \\ -1 - e^{2ik \cdot a_1} & 1 + e^{-2ik \cdot a_3} & 4 \end{pmatrix} \tag{11}$$

The spectrum of $v_k$ is given by

$$\omega_k^2 = \frac{\lambda}{m} \left\{ 1, \frac{5}{2} \pm \frac{1}{2} \sqrt{3 + 2\cos(k_1) + 2\cos(k_1 - k_2) + 2\cos(k_2)} \right\} \tag{12}$$

where $k_1, k_2$ stand for the inner product of k vector with $a_1, a_2$. We can observe the characteristic flat-band and two quadratic bands.

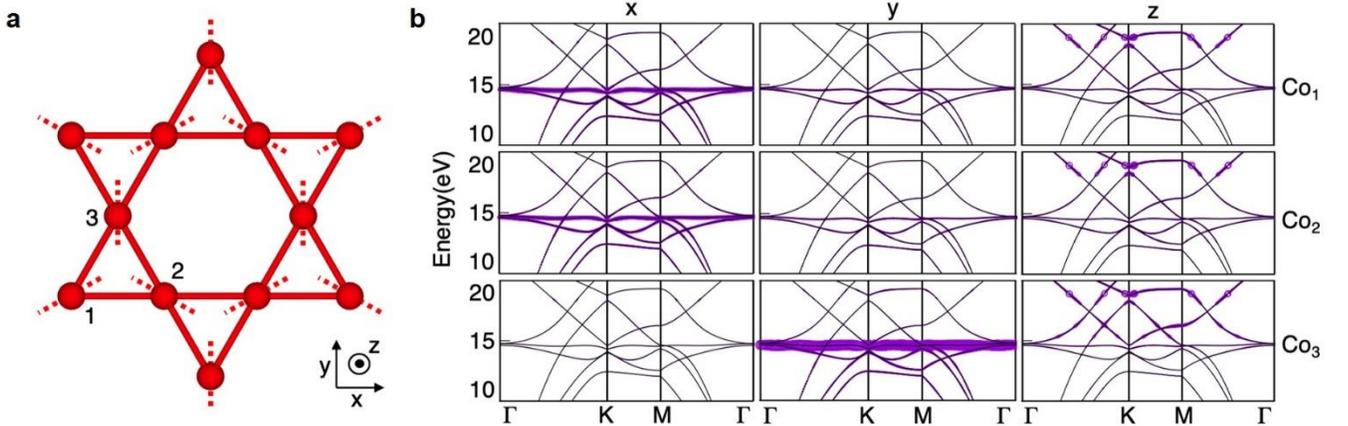

**Fig. 5. Vibration nature of the phonon flat band. a,** Kagome lattice with dotted lines denoting the directions in which the atomic movement is confined. The inner atoms are displaced according to the elementary localized deformation that comprises the flat band. **b,** Atom displacement resolved phonon band structure along x, y and z-direction vibration for $Co_1$, $Co_2$ and $Co_3$ corresponding to Fig. 5a.

**Data availability:** All relevant data are available from the corresponding authors upon reasonable request.

**Acknowledgments:** We acknowledge insightful discussions with Biao Lian and Zhida Song, and technical assistance from Limin Liu and Gaihua Ye. **Funding:** Experimental and theoretical work at Princeton University was supported by the Gordon and Betty Moore Foundation (GBMF4547/ Hasan) and the United States Department of Energy (US DOE) under the Basic Energy Sciences programme (Grant No. DOE/BES DE-FG-02-05ER46200). M.Z.H. acknowledges support from Lawrence Berkeley National Laboratory and the Miller Institute of Basic Research in Science at the University of California, Berkeley in the form of a Visiting Miller Professorship. This work benefited from partial lab infrastructure support under NSF-DMR-1507585. M. Z. H. also acknowledges visiting scientist support from IQIMat the California Institute of Technology. The work at Renmin University was supported by the National Key R&D Program of China (Grants No. 2016YFA0300504 and 2018YFE0202600), the National Natural Science Foundation of China (No. 11774423,11822412), the Fundamental Research Funds for the Central Universities, and the Research Funds of Renmin University of China (RUC) (18XNLG14, 19XNLG17). The authors acknowledge the use of Princeton's Imaging and Analysis Center, which is partially supported by the Princeton Center for Complex Materials, a National Science Foundation (NSF)-MRSEC program (DMR-1420541). T.-R.C. was supported from the Young Scholar Fellowship Program by the Ministry of Science and Technology (MOST) in Taiwan, under MOST Grant for the Columbus Program No. MOST107-2636-M-006-004, National Cheng Kung University, Taiwan, and National Center for Theoretical Sciences (NCTS), Taiwan. Z. W. and K. J. acknowledge US DOE Grant No. DE-FG02-99ER45747. T.N. acknowledges support from the European Union's Horizon 2020 research and innovation programme (ERC-StG-Neupert-757867-PARATOP). The work performed at the Texas Center for Superconductivity at the University of Houston is supported by the U.S. Air Force Office of Scientific Research Grant FA9550-15-1-0236, the T. L. L. Temple Foundation, the John J. and Rebecca